\documentclass[submission,copyright,creativecommons]{eptcs}
\providecommand{\event}{ThEdu 2020} % Name of the event you are submitting to
\usepackage{breakurl}             % Not needed if you use pdflatex only.
\usepackage{underscore}           % Only needed if you use pdflatex.
\usepackage{stmaryrd}
\usepackage{amsfonts}
\usepackage{graphicx}
\usepackage{amsmath}

\usepackage[T1]{fontenc}
\usepackage[utf8x]{inputenc}
\usepackage{amssymb}

\title{Formalizing IMO Problems and Solutions in Isabelle/HOL}
\author{Filip Marić 
\institute{Faculty of Mathematics,}
\institute{University of Belgrade,\\
Belgrade, Serbia}
\email{filip@matf.bg.ac.rs}
\and
Sana Stojanović-Đurđević
\institute{Faculty of Mathematics,}
\institute{University of Belgrade,\\
Belgrade, Serbia}
\email{sana@matf.bg.ac.rs}
}
\def\titlerunning{Formalizing IMO Solutions}
\def\authorrunning{F.~Marić and S.~Stojanović-Đurđević}

\newcommand{\isatext}[1]{\emph{#1}}
\def\isacartoucheopen{\isatext{\raise.3ex\hbox{$\scriptscriptstyle\langle$}}}%
\def\isacartoucheclose{\isatext{\raise.3ex\hbox{$\scriptscriptstyle\rangle$}}}%

\newtheorem{problem}{Problem}

\begin{document}
\maketitle

\begin{abstract}
  The \emph{International Mathematical Olympiad (IMO)} is perhaps the
  most celebrated mental competition in the world and as such is among
  the greatest grand challenges for Artificial Intelligence (AI). The
  \emph{IMO Grand Challenge}, recently formulated, requires to build
  an AI that can win a gold medal in the competition. We present some
  initial steps that could help to tackle this goal by creating a
  public repository of mechanically checked solutions of IMO Problems
  in the interactive theorem prover Isabelle/HOL. This repository is
  actively maintained by students of the Faculty of Mathematics,
  University of Belgrade, Serbia within the course ''Introduction to
  Interactive Theorem Proving''.
\end{abstract}

\section{Introduction}

The International Mathematical Olympiad (IMO) is the World
Championship Mathematics Competition for High School (pre-college)
students and is held annually in a different country. It has gradually
expanded to over 100 countries from 5 continents. The competition
consists of six problems, which pupils solve during two days (each day
they are given three problems and 4.5 hours to work on them). Each
problem is worth seven points for the maximum total score of 42
points. The problems are chosen from various areas of secondary school
mathematics, generally classifiable as geometry, number theory,
algebra, and combinatorics. Problems are expressed with no a priori
knowledge of higher mathematics such as calculus and analysis, and
solutions are often elementary. Few days before the contest, the
International Jury chooses the Olympiad problems. The Jury chooses six
problems from the \emph{Shortlist}, a set of around 30 original,
beautiful, and difficult problems submitted by mathematicians from
around the world. Fully solved shortlists of recent competitions are
available on the official IMO web-site
(\url{http://www.imo-official.org/}), while the solutions of all
shortlisted problems from 1959-2009 are available in \emph{The IMO
  Compendium}~\cite{imocompendium}.

IMO is the oldest of International Science Olympiads and is perhaps
the most known mental competition in the world. As such it is relevant
as one of the greatest grand challenges for Artificial Intelligence
(AI). Recently, a group of scientists gathered around the theorem
prover Lean, has formulated the \emph{IMO Grand
  Challenge}\footnote{\url{http://imo-grand-challenge.github.io/}}:
\textit{Build an AI that can win a gold medal in the competition.} To
remove ambiguity about the scoring rules, the authors of the challenge
propose the formal-to-formal (F2F) variant of the IMO: the AI receives
a formal representation of the problem (in the Lean Theorem Prover),
and needs to emit a formal (i.e. machine-checkable) proof. A
proposal for encoding IMO problems in Lean is currently being
developed.

There are already some automated theorem provers capable of solving
some specific type of problems. For example, algebraic or
semi-algebraic theorem provers are very successful at solving specific
classes of geometry problems. However, such provers rarely produce
formal proofs and they do not offer, synthetic, human-understandable
proofs and justifications (although algebraic proofs can be
mechanically checked by analyzing proof
certificates~\cite{wucoq}). Another example of a successful technique
for automated solving of geometry problems is the so-called \emph{Area
  Method}~\cite{areamethod}. Such methods are usually highly
specialized for specific classes of problems (e.g., only to some
classes of geometric problems), and thus they are very far from
general AI.

In this paper we present formalizations of several IMO problem
solutions, created within the course ``Introduction to interactive
theorem proving'' at the Faculty of Mathematics, University of
Belgrade. Currently problems in algebra, combinatorics and number
theory are formalized (geometry problems are skipped, since their
formalization requires a rich background theory of high-school
synthetic geometry, that is not available in Isabelle/HOL at the
moment). All formalized solutions and problem statements (including
the three solutions presented in this paper) are publicly available in
the GitHub repository \url{http://github.com/filipmaric/IMO}. We hope
that manually constructed proofs and their statements could help
better understand the challenges in formalizing IMO solutions and in
the long run lead to their better automation.

\section{Isabelle/HOL/Isar}

In this section we give a brief overview of terms and notions of the
Isabelle/Isar proof language used in this paper. This will give the
reader only a very rough overview of the syntax used in the rest of
the paper, and we refer him to seek more details in the official
Isabelle/HOL documentation, preferably~\cite{isabelle-prog-prove}.

Isabelle/HOL is an incarnation of a simply typed higher-order logic.
The type system of Isabelle/HOL is very close to functional
programming languages. Base types have the usual names: \emph{bool},
\emph{nat}, \emph{int}, \emph{real}. Type annotations are denoted by
$::$. For example, $3::int$ is an integer constant $3$ (numerals are
supported by default). Type \emph{nat} is an inductive type of natural
numbers in HOL and all values are generated by a constant zero ($0$)
and a constructor ($\mathit{Suc}$). Statements about natural numbers
are usually proved by induction. Set types are also supported. The
type $'a\ set$ denotes the type of sets with elements of type $'a$
(e.g., $nat\ set$ denotes sets of natural numbers). Set of natural
numbers less than $n$, $\{0,1,\ldots,n-1\}$, is denoted by {\it
  $\{0..$<$n\}$}; set of natural numbers less than or equal to $n$,
$\{0,1,\ldots,n\}$, is denoted by {\it $\{0..n\}$}; set of natural
numbers greater than $n$ is denoted by {\it \{n+1..\}}. Function types
are denoted by $'a \Rightarrow 'b$. For example, function $f$ from
integer to real numbers is denoted by $f :: nat \Rightarrow int$.

Terms are built from variables and constants by applying functions and
operators. Functions are curried and function application is written
in prefix form (as in most functional languages). Terms can also use
some advanced mathematical notation. For example {\it $\sum$ k=0..n. f
  k} denotes the sum $f(0)+f(1)+\ldots+f(n)$. The same sum can also be
denoted by {\it $\sum$ k$\leq$ n. f k}. Formulae
(terms of type \emph{bool}) are built using the usual boolean
connectives ($\wedge$, $\vee$, $\neg$, $\longrightarrow$,
$\longleftrightarrow$), and quantifiers $\forall x.\ P\ x$ and
$\exists x.\ P\ x$. Isabelle/HOL also supports the indefinite
description operator {\it SOME x. P x}, describing any element that
satisfies the property {\it P} (assuming such element exists) and the
definite description operator {\it THE x. P x} describing the unique
element that satisfies the property {\it P} (assuming such element
exists).

The language Isar is used for writing readable, textbook-like
structured theories and proofs. Definitions are introduced by the
keyword {\bf definition}, followed by the name of the constant or the
function being defined, its type and a defining equality. Recursive
functions are defined using the keywords {\bf primrec} and {\bf
  fun}. Keywords {\bf lemma} and {\bf theorem} are used to state the
statement being proved. Statements usually have the form:

{\it
  \begin{tabbing}
    \hspace{5mm}\=\hspace{5mm}\=\hspace{5mm}\=\kill
    {\bf lemma}\\
    \>{\bf fixes} $\isacartoucheopen$variables$\isacartoucheclose$\\
    \>{\bf assumes} $\isacartoucheopen$assumptions$\isacartoucheclose$\\
    \>{\bf shows}  $\isacartoucheopen$goal$\isacartoucheclose$
  \end{tabbing}
}

Types of variables used in the statement (or in the proof) can be
stated upfront with the keyword {\bf fixes} (Isabelle supports
type-inference so variables need not always be explicitly
declared). Assumptions can be stated after the {\bf assumes} keyword,
and conclusions must be stated after the {\bf shows} keyword.

Lemmas and proof rules can also be specified in Isabelle's meta-logic
syntax:

{\it
  \begin{tabbing}
    \hspace{5mm}\=\hspace{5mm}\=\hspace{5mm}\=\kill
    $\bigwedge$ x$_1$, \ldots, x$_n$. $\llbracket$ assumption$_1$, \ldots, assumption$_k$ $\rrbracket$ $\Longrightarrow$ goal
  \end{tabbing}
}

Each lemma and theorem must be followed by a proof. Proofs can be
either automatic or interactive, structured proofs.

Automated proofs are specified by a keyword {\bf by} followed by the
name of the automated proof method used (most often these are {\it
  simp}, {\it auto}, {\it force}, {\it blast}, {\it metis}, {\it smt},
{\it presburger}) and possibly by additional parameters.

Structured proofs in Isar are written within a {\bf proof}-{\bf qed}
block. Opening keyword {\bf proof} can be followed by the proof
method that is applied in the beginning of the proof. Proofs by case
analysis are specified using the method {\it cases} (e.g., {\bf proof}
{\it (cases "n > 0")}). Proofs by induction are specified using the
method {\it induction} (e.g., {\bf proof} {\it (induction n rule:
  less\_induct)}). If method is not specified, the system automatically
chooses the first method (to be applied). If the proof starts by {\bf
  proof}{\it -}, then no method is being applied.

A typical proof introduces a chain of intermediate
statements. Intermediate statements in the proof are given using the
keyword {\bf have}, and the final statement (the goal of the current
proof) is given using the keyword {\bf show}. This gives the following
proof structure.

{\it
  \begin{tabbing}
    \hspace{5mm}\=\hspace{5mm}\=\hspace{5mm}\=\kill
    {\bf proof}-\\
    \> {\bf have} "statement$_1$" $\isacartoucheopen$proof$\isacartoucheclose$\\
    \> {\bf have} "statement$_2$" $\isacartoucheopen$proof$\isacartoucheclose$\\
    \> ...\\
    \> {\bf have} "statement$_k$" $\isacartoucheopen$proof$\isacartoucheclose$\\
    \> {\bf show} ?thesis $\isacartoucheopen$proof$\isacartoucheclose$\\
    {\bf qed}
  \end{tabbing}
}

\noindent Here {\it ?thesis} abbreviate the goal of the current
proof-block. Each statement introduced by key word {\bf have} or {\bf show}
must have its own proof (that, again can be specified either using
{\bf by} or {\bf proof}-{\bf qed} block). If automated provers
(invoked by the keyword {\bf by}) need to use additional facts,
those facts must be explicitly ''pumped into the proof
context''. There are many ways this can be done (for simplicity, we
shall use only few). The simplest one is to use the keyword {\bf
  using} after the statement or the keyword {\bf from} before the
statement, followed by the fact that is being inserted into the proof
context (and therefore made available to the automated prover). If
needed, facts can be named (just beforehand they are stated) and those
names can be used instead of explicitly writing the fact enclosed by
cartouches $\isacartoucheopen\ldots\isacartoucheclose$. If the facts
that are used are part of the assumptions of the current lemma, they
can be accessed using the abbreviation {\it assms}.

Usually, intermediate statements are chained and the next statement is
proved using the previous one. This gives the following proof structure.

{\it
  \begin{tabbing}
    \hspace{5mm}\=\hspace{5mm}\=\hspace{5mm}\=\kill
    {\bf proof}-\\
    \> {\bf have} "statement$_1$" $\isacartoucheopen$proof$\isacartoucheclose$\\
    \> {\bf then have} "statement$_2$" $\isacartoucheopen$proof$\isacartoucheclose$\\
    \> ...\\
    \> {\bf then show} ?thesis $\isacartoucheopen$proof$\isacartoucheclose$\\
    {\bf qed}
  \end{tabbing}
}

Another often encountered proof structure is when several intermediate
statements are used to prove the final goal. In Isar, this can be
specified using the combination {\bf moreover}-{\bf ultimately}.

{\it
  \begin{tabbing}
    \hspace{5mm}\=\hspace{5mm}\=\hspace{5mm}\=\kill
    {\bf proof}-\\
    \> {\bf have} "statement$_1$" $\isacartoucheopen$proof$\isacartoucheclose$\\
    \> {\bf moreover have} "statement$_2$" $\isacartoucheopen$proof$\isacartoucheclose$\\
    \> ...\\
    \> {\bf moreover have} "statement$_k$" $\isacartoucheopen$proof$\isacartoucheclose$\\
    \> {\bf ultimately show} ?thesis $\isacartoucheopen$proof$\isacartoucheclose$\\
    {\bf qed}
  \end{tabbing}
}

Proofs often include chains of equations or inequalities. These are
supported in Isar using keywords {\bf also}-{\bf finally}.

{\it
  \begin{tabbing}
    \hspace{5mm}\=\hspace{5mm}\=\hspace{5mm}\=\kill
    {\bf proof}-\\
    \> {\bf have} "t$_1$ = t$_2$" $\isacartoucheopen$proof$\isacartoucheclose$\\
    \> {\bf also have} "... = t$_3$" $\isacartoucheopen$proof$\isacartoucheclose$\\
    \> ...\\
    \> {\bf also have} "... = t$_k$" $\isacartoucheopen$proof$\isacartoucheclose$\\
    \> {\bf finally show} ?thesis $\isacartoucheopen$proof$\isacartoucheclose$\\
    {\bf qed}
  \end{tabbing}
}

The last proof, after the keyword {\bf finally} has the fact
$t_1 = t_k$ pumped into its proof context. Instead of a chain of
equalities, a chain of inequalities or a mixed chain of equalities and
inequalities can be used.

\section{Example Problems}

In this section we shall describe formalizations of three
characteristic problems --- one in algebra, one in combinatorics and
one in number theory.

\subsection{A Problem in Algebra}

As an example of an algebraic proof that is very straightforward to
formalize, we will describe the problem A1 from 2006\footnote{The
  problem statement and its solution are described in the official
  competition bulletin
  \url{http://www.imo-official.org/problems/IMO2006SL.pdf}.}. Since
this proof is very short (both in informal and in formal language), we
shall show it in much detail.

We will first present the official solution to this problem, and then
we will analyze the formulation of this problem in Isabelle/Isar form:
\\

%\noindent
\fbox{
\begin{minipage}{0.8\textwidth}
{\footnotesize
    
\begin{problem}[2006 A2]
  The sequence of real numbers $a_0$, $a_1$, $a_2$, \ldots is defined
  recursively by

  $$a_0 = -1, \qquad \sum_{k=0}^{n}\frac{a_{n-k}}{k+1} = 0 \quad \textrm{for}\quad n \geq 1.$$

\noindent Show that $a_n > 0$ for $n \geq 1$.
\end{problem}

}
\end{minipage}
}

%\noindent
\fbox{
\begin{minipage}{0.8\textwidth}
{\footnotesize

\textbf{Solution}. The proof goes by induction. For $n=1$ the formula yields $a_1 =
1/2$. Take $n \geq 1$, assume $a_1, \ldots, a_n > 0$ and write the
recurrence formula for $n$ and $n+1$, respectively as

$$\sum_{k=0}^n \frac{a_k}{n - k + 1} = 0\qquad \textrm{and}\qquad \sum_{k=0}^{n+1} \frac{a_k}{n-k+2} = 0.$$

\noindent Subtraction yields

\begin{align*}
0 & = (n+2)\sum_{k=0}^{n+1} \frac{a_k}{n-k+2} - (n+1)\sum_{k=0}^n \frac{a_k}{n - k + 1} = (n + 2)a_{n+1} + \sum_{k=0}^{n} \left(\frac{n + 2}{n - k + 2} - \frac{n + 1}{n - k + 1}\right) a_k.
\end{align*}
  
The coefficient of $a_0$ vanishes, so
$$a_{n+1} = \frac{1}{n+2}\sum_{k=1}^n\left(\frac{n+1}{n-k+1} - \frac{n+2}{n-k+2}\right)a_k = \frac{1}{n+2}\sum_{k=1}^n \frac{k}{(n-k+1)(n-k+2)} a_k.$$

The coefficients of $a_1, \ldots, a_n$ are all positive. Therefore,
$a_1, \ldots, a_n > 0$ implies $a_{n+1} > 0$.
}
\end{minipage}
}

\vspace{1em}
Writing the statement in the formal language is very straightforward
(the sequence $a_n$ is modeled by a function that maps natural indices
to real values).

{\it
\begin{tabbing}
\hspace{5mm}\=\hspace{5mm}\=\hspace{5mm}\=\kill
{\bf theorem} IMO_2006_SL_A2:\\
\>  {\bf fixes} a :: "nat $\Rightarrow$ real"\\
\>  {\bf assumes} "a 0 = -1" "$\forall$ n $\geq$ 1. ($\sum$ k $\leq$ n. a (n - k) / (k + 1)) = 0"\\
\>  {\bf assumes} "n $\geq$ 1"\\
\>  {\bf shows} "a n $>$ 0"
\end{tabbing}
}

\noindent Note that the application of the division operator
implicitly casts the natural number $k + 1$ to real.

In the official solution it is stated that induction is used, but it
is not explicitly stated what induction principle is used. A careful
examination of the official proof reveals that complete (strong)
induction must be used. In the official proof, it is shown that
$a_{n+1} > 0$ holds, assuming that $a_k > 0$ holds for all $1 \leq k
\leq n$. In Isabelle/HOL, the strong induction principle is given by
the rule {\it less\_induct}. All variables are universally
quantified.\footnote{A more faithful formulation of the {\it
    less\_induct} rule would be {\it ($\bigwedge$ x. ($\bigwedge$ y. y
    $<$ x $\Longrightarrow$ ?P y) $\Longrightarrow$ ?P x)
    $\Longrightarrow$ ?P ?a}. In Isabelle, once the theorem is proved,
  the object logic variables are lifted into schematic variables. For
  readability reasons, in this paper we will assume implicit universal
  quantification.}

{\it
  \begin{tabbing}
($\bigwedge$ x. ($\bigwedge$ y. y $<$ x $\Longrightarrow$ P y) $\Longrightarrow$ P x) $\Longrightarrow$ P a
\end{tabbing}
}

\noindent This rule is used in the proof of the theorem.

{\it
  \begin{tabbing}
\hspace{5mm}\=\hspace{5mm}\=\kill
\ldots\\
\>{\bf shows} "a n $>$ 0"\\
\> {\bf using} \isacartoucheopen n $\geq$ 1\isacartoucheclose\\
{\bf proof} (induction n rule: less_induct)\\
\>  {\bf case} (less n)\\
\>  {\bf show} ?case\\
\>\>     ...\\
{\bf qed}
\end{tabbing}
}

In formal proof, to show that {\it a n $>$ 0} holds for arbitrary
$n \geq 1$, it suffices to show that {\it a n $>$ 0} holds for
arbitrary $n \geq 1$, under the assumption that for any $1 \leq k < n$
it holds that {\it a k $>$ 0}. Therefore, many indices in our formal
proof will be shifted by one from the indices in the official,
informal proof (since in formal proof we show that {\it a n $>$ 0},
while informal proof shows that $a_{n+1} > 0$).

The proof then distinguishes the base case ($n = 1$) and the inductive
step (when $n > 1$).

{\it
\begin{tabbing}
\hspace{5mm}\=\hspace{5mm}\=\hspace{5mm}\=\kill
\>{\bf show} ?case\\
\>{\bf proof} cases\\
\>\>{\bf assume} n $=$ 1\\
\>\>\ldots\\
\>{\bf next}\\
\>\>{\bf assume} n $\neq$ 1\\
\>\>\ldots\\
\>{\bf qed}
\end{tabbing}
}

The induction base is quite easily discharged. The informal proof just
states ''For $n = 1$ the formula yields $a_1 = 1/2$'', and this is
quite directly formalized.

{\it
\begin{tabbing}
  \hspace{5mm}\=\hspace{5mm}\=\hspace{5mm}\=\kill
  \>{\bf assume} "n = 1"\\
  \>{\bf have} "a 1 = 1/2" {\bf using} assms {\bf by} auto\\
  \>{\bf then show} ?thesis {\bf using} \isacartoucheopen n = 1\isacartoucheclose\ {\bf by} simp
\end{tabbing}
}

Note that {\it a 1 = 1/2} follows automatically from the theorem
assumptions ($a_0 = -1$ and $\sum_{k=0}^{n}\frac{a_{n-k}}{k+1} = 0$,
denoted by {\it assms}), and this shows the goal {\it a n $>$ 0}
(denoted by {\it ?thesis}), since in this case it holds that $n = 1$.

As usual, the inductive step is much harder. The informal proof goes
as follows. ''Take $n \geq 1$, assume $a_1, \ldots, a_n > 0$ and write the
recurrence formula for $n$ and $n+1$, respectively as

$$\sum_{k=0}^n \frac{a_k}{n - k + 1} = 0\qquad \textrm{and}\qquad \sum_{k=0}^{n+1} \frac{a_k}{n-k+2} = 0.$$

\noindent Subtraction yields

$$0 = (n+2) \sum_{k=0}^{n+1} \frac{a_k}{n-k+2} - (n+1) \sum_{k=0}^n \frac{a_k}{n - k + 1}.\textrm{''}$$

Formalization follows this quite faithfully (except that indices are
shifted by one).

{\it
\begin{tabbing}
\hspace{5mm}\=\hspace{5mm}\=\hspace{5mm}\=\kill
\>      {\bf have} "($\sum$ k $<$ n. a k / (n - k)) = 0"\\
\>\>        {\bf using} assms(2)[of "n - 1"] \isacartoucheopen n $>$ 1\isacartoucheclose\ sum.nat_diff_reindex[of "$\lambda$ k. a k / (n - k)" "n"] \\
\>\>        {\bf by} simp\\
\>      {\bf moreover} {\bf have} "($\sum$ k $<$ n + 1. a k / (n + 1 - k)) = 0"\\
\>\>        {\bf using} assms(2)[of "n"] \isacartoucheopen n $>$ 1\isacartoucheclose\ sum.nat_diff_reindex[of "$\lambda$ k. a k / (n + 1 - k)" "n + 1"]\\
\>\>        {\bf by} simp\\
\>      {\bf ultimately} {\bf have} "(n + 1) * ($\sum$ k $<$ n + 1. a k / (n + 1 - k)) - n * ($\sum$ k $<$ n. a k / (n - k)) = 0"\\
\>\> {\bf by} simp
\end{tabbing}
}

The proof steps use the lemma {\it sum.nat\_diff\_reindex}, already
available in Isabelle/HOL:

{\it
\begin{tabbing}
\hspace{5mm}\=\hspace{5mm}\=\hspace{5mm}\=\kill
($\sum$ i $<$ n. g (n - Suc i)) = ($\sum$ $i$ $<$ n. g i)
\end{tabbing}
}

\noindent It's use in the informal proof is only implicit (sum
re-indexing is considered fully trivial).

The informal proof continues by giving the equality between the
difference of two sums with the expression:

$$(n + 2)a_{n+1} + \sum_{k=0}^{n}\left(\frac{n + 2}{n - k + 2} - \frac{n + 1}{n - k + 1}\right) a_k.$$

This is also done in the formal proof, except that indices are again
off by one.

{\it
\begin{tabbing}
  \hspace{5mm}\=\hspace{5mm}\=\hspace{5mm}\=\kill
 {\bf then} {\bf have} "(n + 1) * a n = - ($\sum$ k $<$ n. ((n + 1) / (n + 1 - k) - n / (n - k)) * a k)"\\
\>   {\bf by} (simp add: algebra_simps sum_distrib_left sum_subtractf)\\
 {\bf then} {\bf have} "(n + 1) * a n = ($\sum$ k $<$ n. (n / (n - k) - (n + 1) / (n + 1 - k)) * a k)"\\
\>   {\bf by} (simp add: algebra_simps sum_negf[symmetric])
\end{tabbing}
}

The first proof step uses various algebraic simplifications (contained
in the collection of theorems called {\it algebra\_simps}) and the
following two properties of sums ({\it sum\_distrib\_left} and {\it
  sum\_subtractf}):

{\it
\begin{tabbing}
\hspace{5mm}\=\hspace{5mm}\=\hspace{5mm}\=\kill
r * ($\sum$ n $\in$ A. f n) = ($\sum$n$\in$A. r * f n)\qquad\qquad ($\sum$x$\in$A. f x - g x) = ($\sum$x$\in$A. f x) - ($\sum$x$\in$A. g x)
\end{tabbing}
}

The second proof step uses various algebraic simplifications
(collection of theorems {\it algebra\_simps}) and the following
property of sums ({\it sum\_negf}):

{\it
\begin{tabbing}
\hspace{5mm}\=\hspace{5mm}\=\hspace{5mm}\=\kill
($\sum$x$\in$A. - f x) = - ($\sum$x$\in$A. f x)
\end{tabbing}
}

Note that although this formal proof is essentially the same as the
informal proof, it is a bit more verbose. Some intermediate steps had
to be specified and proved using several lemmas already available in
Isabelle/HOL. Without the use of intermediate steps, automated provers
in Isabelle were not able to directly prove the final goal.

The informal proof continues: ''The coefficient of $a_0$ vanishes, so

$$a_{n+1} = \frac{1}{n+2}\sum_{k=1}^n\left(\frac{n+1}{n-k+1} - \frac{n+2}{n-k+2}\right)a_k = \frac{1}{n+2}\sum_{k=1}^n \frac{k}{(n-k+1)(n-k+2)} a_k.$$

The coefficients of $a_1, \ldots, a_n$ are all positive. Therefore,
$a_1, \ldots, a_n > 0$ implies $a_{n+1} > 0$''.

The formal proof roughly follows this. 

{\it
\begin{tabbing}
\hspace{5mm}\=\hspace{5mm}\=\hspace{5mm}\=\kill
{\bf also} {\bf have} "... = ($\sum$ k $\in$ $\{$1..$<$n$\}$. (n / (n - k) - (n + 1) / (n + 1 - k)) * a k)"\\
\> {\bf using} \isacartoucheopen n $>$ 1\isacartoucheclose \\
\> {\bf by} (subst sum_remove_zero, auto)
\end{tabbing}
}

It is easily automatically deduced that the coefficient of $a_0$ is $n
/ (n - 0) - (n + 1) / (n + 1 - 0) = 0$. However, it was necessary to formulate, and to prove, a separate lemma for isolating the first member of the sum.

{\it
\begin{tabbing}
  \hspace{5mm}\=\hspace{5mm}\=\hspace{5mm}\=\kill
{\bf lemma} sum_remove_zero:\\
\>  {\bf fixes} n :: nat\\
\>  {\bf assumes} "n $>$ 0"\\
\>  {\bf shows} "($\sum$ k $<$ n. f k) = f 0 + ($\sum$ k $\in$ $\{$1..$<$n$\}$. f k)"\\
\>  {\bf using} assms\\
\>  {\bf by} (simp add: atLeast1_lessThan_eq_remove0 sum.remove)
\end{tabbing}
}

Next, in the informal proof it trivially holds that the sum is positive,
since the coefficients of $a_1, \ldots, a_n$ are all positive. But in
the formal proof it is proved that the sum is positive by proving
that all its members are positive. This is given by the Isabelle/HOL
rule {\it sum\_pos}.

{\it
\begin{tabbing}
  \hspace{5mm}\=\hspace{5mm}\=\hspace{5mm}\=\kill
$\llbracket$finite I; I $\neq$ $\{\}$; $\bigwedge$ i. i $\in$ I $\Longrightarrow$ 0 $<$ f i$\rrbracket$ $\Longrightarrow$ 0 $<$ ($\sum$ x $\in$ I. f x)
\end{tabbing}
}

Applying this rule requires proving it's three assumptions (separated
by the word \emph{next}).

{\it
\begin{tabbing}
  \hspace{5mm}\=\hspace{5mm}\=\hspace{5mm}\=\hspace{5mm}\=\kill
    {\bf also} {\bf have} "... $>$ 0"\\
    {\bf proof} (rule sum_pos)\\
\>      {\bf show} "finite $\{$1..$<$n$\}$" {\bf by} simp\\
    {\bf next}\\
\>      {\bf show} "$\{$1..$<$n$\}$ $\neq$ \{\}" {\bf using} \isacartoucheopen n $>$ 1\isacartoucheclose\ {\bf by} simp\\
    {\bf next}\\
\>    {\bf fix} i\\
\>    {\bf assume} "i $\in$ $\{$1..$<$n$\}$"\\
\>    {\bf show} "(n / (n - i) - (n + 1) / (n + 1 - i)) * a i $>$ 0" ({\bf is} "?ci * a i $>$ 0")\\
\>    {\bf proof}-\\
\>\>      {\bf have} "a i $>$ 0" {\bf using} less  \isacartoucheopen i $\in$ $\{$1..$<$n$\}$ \isacartoucheclose\ {\bf by} simp\\
\>\>      {\bf moreover} {\bf have} "?ci $>$ 0"\\
\>\>      {\bf proof}-\\
\>\>\>          {\bf have} "?ci = i / ((n - i) * (n + 1 - i))" {\bf using} \isacartoucheopen i $\in$ $\{$1..$<$n$\}$\isacartoucheclose\ {\bf by} (simp add: field_simps of_nat_diff)\\
\>\>\>          {\bf then} {\bf show} ?thesis {\bf using} \isacartoucheopen i $\in$ $\{$1..$<$n$\}$\isacartoucheclose\  {\bf by} simp\\
\>\>      {\bf qed}\\
\>\>      {\bf ultimately} {\bf show} ?thesis {\bf by} simp\\
{\bf qed}
\end{tabbing}
}

The only non-trivial part in this proof is proving that the
coefficient {\it ?ci} of {\it a i} is non-negative (the first step of
the third assumption). This is done essentially in the same way as in
the informal proof --- two fractions are subtracted, and reduced to a
common fraction with a numerator and denominator that are obviously
positive. To show that each {\it a i} is positive, the induction
hypothesis (denoted by {\it less}) is used.

The proof finishes by noting that we have proved that {\it (n+1)
  $\cdot$ (a n)} is positive, and that, since {\it n+1} is positive,
so must also be {\it a n} (the last proof is found by the Sledgehammer
tool and uses an SMT solver). This step is implicit in informal proof.

{\it
\begin{tabbing}
  \hspace{5mm}\=\hspace{5mm}\=\hspace{5mm}\=\hspace{5mm}\=\kill
  {\bf finally} {\bf have} "(n + 1) * (a n) $>$ 0" .\\
  {\bf then} {\bf show} ?thesis {\bf by} (smt mult_nonneg_nonpos of_nat_0_le_iff)
\end{tabbing}
}

\subsection{A Problem in Combinatorics}

As an example of a problem that has a very short and elegant informal
solution, but which is hard to formalize we show the problem C1 from
2017\footnote{The problem statement and its solution are described in
  the official competition bulletin
  \url{http://www.imo-official.org/problems/IMO2017SL.pdf}.}.

\begin{problem}[2017 C1]
  A rectangle $\mathcal{R}$ with odd integer side lengths is divided into small
  rectangles with integer side lengths. Prove that there is at least
  one among the small rectangles whose distances from the four sides of
  $\mathcal{R}$ are either all odd or all even.
\end{problem}

\begin{figure}[ht]
  \centering

  \includegraphics{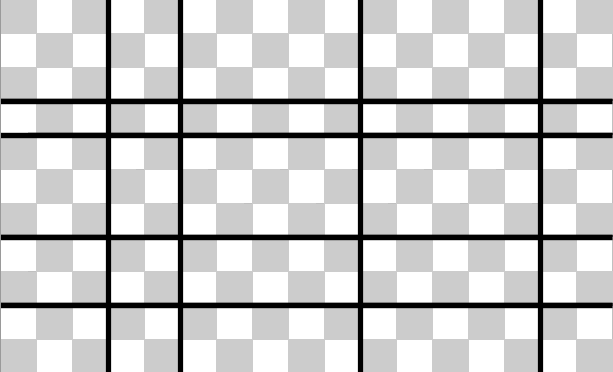}
  \hspace{1cm}
  \includegraphics{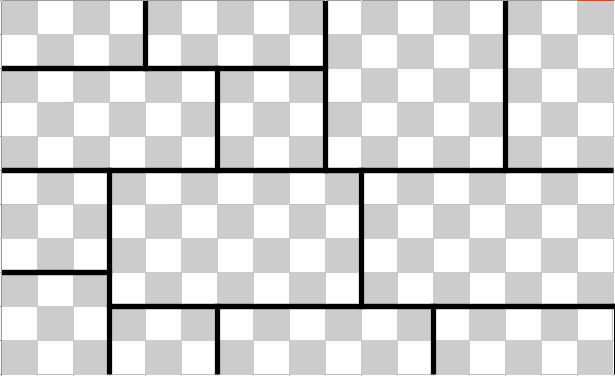}

  \caption{Tiling of a rectangle}
  \label{fig:rect}
\end{figure}

The first major challenge is to give a formal statement of the
problem. Although at first glance one might think that division must
form a rectangular grid (as shown on the left picture in Figure
\ref{fig:rect}), more general tilings are allowed (as shown on the
right picture in Figure \ref{fig:rect}).

We shall assume that a coordinate system is introduced so that the
lower left corner of the big rectangle is in its origin. Each
rectangle will be determined by four non-negative integers $(x_1, x_2,
y_1, y_2)$: coordinates of its left, right, bottom and top line. Unit
squares are indicated by checkerboard pattern, so our big rectangle
can be represented by the quadruple $(0, 17, 0, 11)$. For a rectangle
to be valid (non-empty) it must hold that $x_1 < x_2$ and that $y_1 <
y_2$.

We formalize this as follows (since all coordinates are positive,
instead of integers we use natural numbers). The validity condition
could have been encoded in the rectangle type, but that would require
using a bit more advanced features of Isabelle/HOL, so we did not go
in that direction.

{\it
\begin{tabbing}
  \hspace{5mm}\=\hspace{5mm}\=\hspace{5mm}\=\hspace{5mm}\=\kill
{\bf type\_synonym} rect = "nat $\times$ nat $\times$ nat $\times$ nat"\\[1ex]
{\bf fun} valid\_rect :: "rect $\Rightarrow$ bool" {\bf where} "valid\_rect (x$_1$, x$_2$, y$_1$, y$_2$) $\longleftrightarrow$ x$_1$ $<$ x$_2$ $\wedge$ y$_1$ $<$ y$_2$"
\end{tabbing}
}

Each unit square in a rectangle is characterized by its two integer
coordinates (of its lower left corner). Each valid rectangle contains
a set of unit squares that can be obtained by their coordinates as a
Cartesian product of two discrete integer intervals.

{\it
\begin{tabbing}
  \hspace{5mm}\=\hspace{5mm}\=\hspace{5mm}\=\hspace{5mm}\=\kill
  {\bf type\_synonym} square = "nat $\times$ nat"\\
{\bf fun} squares :: "rect $\Rightarrow$ square set" {\bf where} "squares (x$_1$, x$_2$, y$_1$, y$_2$) = $\{$x$_1$..$<$x$_2$$\}$ $\times$ $\{$y$_1$..$<$y$_2$$\}$"
\end{tabbing}
}

We define a tiling (a subdivision) of a rectangle $\mathcal{R}$ to be
a set of non-overlapping rectangles that cover $\mathcal{R}$. Two
rectangles overlap if they share a common square. A set of rectangles
is non-overlapping if no two different rectangles overlap. A set of
rectangles cover a given rectangle $\mathcal{R}$ if the union of all
squares is equal to the set of squares of $\mathcal{R}$. We formalize
this as follows.

{\it
\begin{tabbing}
  \hspace{5mm}\=\hspace{5mm}\=\hspace{5mm}\=\hspace{5mm}\=\kill
{\bf definition} overlap :: "rect $\Rightarrow$ rect $\Rightarrow$ bool" {\bf where} "overlap r$_1$ r$_2$ $\longleftrightarrow$ squares r$_1$ $\cap$ squares r$_2$ $\neq$ $\{\}$"\\[1ex]
{\bf definition} non\_overlapping :: "rect set $\Rightarrow$ bool" {\bf where}\\
\>  "non_overlapping rs $\longleftrightarrow$ ($\forall$ r$_1$ $\in$ rs. $\forall$ r$_2$ $\in$ rs. r$_1$ $\neq$ r$_2$ $\longrightarrow$ $\neg$ overlap r$_1$ r$_2$)"\\[1ex]
{\bf definition} cover :: "rect set $\Rightarrow$ rect $\Rightarrow$ bool" {\bf where} "cover rs r $\longleftrightarrow$ ($\bigcup$ (squares ` rs)) = squares r"\\[1ex]
{\bf definition} tiles :: "rect set $\Rightarrow$ rect $\Rightarrow$ bool" {\bf where} "tiles rs r $\longleftrightarrow$ cover rs r $\wedge$ non_overlapping rs"
\end{tabbing}
}

Finally, we can give the formal statement of the theorem.

{\it
\begin{tabbing}
  \hspace{5mm}\=\hspace{5mm}\=\hspace{5mm}\=\hspace{5mm}\=\kill
  {\bf theorem} IMO_2017_SL_C1:\\
\>  {\bf fixes} a b :: nat\\
\>  {\bf assumes} "odd a" "odd b" "tiles rs (0, a, 0, b)" "$\forall$ r $\in$ rs. valid\_rect r"\\
\>  {\bf shows} "$\exists$ (x$_1$, x$_2$, y$_1$, y$_2$) $\in$ rs. {\rm l}\={\rm et} ds = $\{$x$_1$ - 0, a - x$_2$, y$_1$ - 0, b - y$_2$$\}$\\
\>\>                   {\rm in} ($\forall$ d $\in$ ds. even d) $\vee$ ($\forall$ d $\in$ ds. odd d)"
\end{tabbing}
}

The informal proof is very short and elegant.

''Let the width and height of $\mathcal{R}$ be odd numbers $a$ and
$b$. Divide $\mathcal{R}$ into $ab$ unit squares and color them green
and yellow in a checkered pattern. Since the side lengths of $a$ and $b$
are odd, the corner squares of $\mathcal{R}$ will all have the same
color, say green. Call a rectangle (either $\mathcal{R}$ or a small
rectangle) green if its corners are all green; call it yellow if the
corners are all yellow, and call it mixed if it has both green and
yellow corners. In particular, $\mathcal{R}$ is a green rectangle.''

Several definitions are introduced to formalize this passage, and one
simple lemma is proved.

{\it
\begin{tabbing}
  \hspace{5mm}\=\hspace{5mm}\=\hspace{5mm}\=\hspace{5mm}\=\kill
  {\bf fun} green :: "square $\Rightarrow$ bool" {\bf where} "green (x, y) $\longleftrightarrow$ (x + y) mod 2 = 0"\\
  {\bf fun} yellow :: "square $\Rightarrow$ bool" {\bf where} "yellow (x, y) $\longleftrightarrow$ (x + y) mod 2 $\neq$ 0"\\[1ex]
  {\bf fun} corners :: "rect $\Rightarrow$ square set" {\bf where}\\
\>  "corners (x$_1$, x$_2$, y$_1$, y$_2$) = $\{$(x$_1$, y$_1$), (x$_1$, y$_2$-1), (x$_2$-1, y$_1$), (x$_2$-1, y$_2$-1)$\}$"\\[1ex]
  {\bf definition} green\_rect :: "rect $\Rightarrow$ bool" {\bf where} "green\_rect r $\longleftrightarrow$ ($\forall$ c $\in$ corners r. green c)"\\
  {\bf definition} yellow\_rect :: "rect $\Rightarrow$ bool" {\bf where} "yellow\_rect r $\longleftrightarrow$ ($\forall$ c $\in$ corners r. yellow c)"\\
  {\bf definition} mixed\_rect ::  "rect $\Rightarrow$ bool" {\bf where} "mixed\_rect r $\longleftrightarrow$ $\neg$ green\_rect r $\wedge$ $\neg$ yellow\_rect r"\\[1ex]
{\bf lemma}\\
\>{\bf assumes} "odd a" "odd b"\\
\>{\bf shows} "green_rect (0, a, 0, b)"\\
\>{\bf unfolding} green\_rect\_def {\bf by} auto
\end{tabbing}
}

The informal proof continues as follows. ''We will use the following trivial observations.
\begin{itemize}
\item Every mixed rectangle contains the same number of green and yellow squares;
\item Every green rectangle contains one more green square than yellow square;
\item Every yellow rectangle contains one more yellow square than green square.''
\end{itemize}

This is where things start to get harder. Unfortunately, these
observations seem far from trivial, when it comes to their formal
proofs. Giving their formal statement requires following two definitions.

{\it
\begin{tabbing}
  \hspace{5mm}\=\hspace{5mm}\=\hspace{5mm}\=\hspace{5mm}\=\kill
{\bf definition} green_squares :: "rect $\Rightarrow$ square set" {\bf where}\\
\>   "green_squares r = $\{$(x, y) $\in$ squares r. green (x, y)$\}$"\\
{\bf definition} yellow_squares :: "rect $\Rightarrow$ square set" {\bf where}\\
\>   "yellow_squares r = $\{$(x, y) $\in$ squares r. yellow (x, y)$\}$"
\end{tabbing}
}

Stating the observations is now straightforward (we only show the
middle one).

{\it
\begin{tabbing}
  \hspace{5mm}\=\hspace{5mm}\=\hspace{5mm}\=\hspace{5mm}\=\kill
{\bf lemma} green_rect: \\
\>  {\bf assumes} "valid_rect (x$_1$, x$_2$, y$_1$, y$_2$)"  "green_rect (x$_1$, 
x$_2$, y$_1$, y$_2$)"\\
\>  {\bf shows} "card (green_squares (x$_1$, x$_2$, y$_1$, y$_2$)) = card (yellow_squares (x$_1$, x$_2$, y$_1$, y$_2$)) + 1"
\end{tabbing}
}

There are two main approaches to prove this lemma. The first approach
requires explicitly calculating the number of green and the number of
yellow squares in a green rectangle. The other approach requires to
establish a bijective mapping between all yellow squares and all but
one green square in a rectangle. We have taken the first approach. The
number of green and yellow squares in a rectangle depends on the
rectangle dimensions, but also on whether the first square (the one
with the coordinates $(0, 0)$) is a green or yellow square. We will
divide the set of squares (in a rectangle) into two halves in the
following manner. Denote the total number of squares by $k$. If $k$ is
even, those halves will be equal, and if $k$ is odd, one half will be
greater by one square than the other half. In both cases, the number
of squares of the half that contains the starting square is equal
$k/2$ rounded upwards (by the ceiling function) and the number of
squares of the other half is equal $k/2$ rounded downwards (by the
floor function). This is formalized by the following lemmas (the
expression {\it k {\rm div} 2} rounds $k/2$ downwards, and {\it (k +
  1) {\rm div} 2} rounds it upwards).

{\it
\begin{tabbing}
  \hspace{5mm}\=\hspace{5mm}\=\hspace{5mm}\=\hspace{5mm}\=\kill
{\bf lemma}\\
\>  {\bf assumes} "green (x$_1$, y$_1$)" "valid\_rect (x$_1$, x$_2$, y$_1$, y$_2$)"\\
\>  {\bf shows} \="card (yellow_squares (x$_1$, x$_2$, y$_1$, y$_2$)) = ((x$_2$ - x$_1$) * (y$_2$ - y$_1$)) {\rm div} 2"\\
\>\>        "card (green_squares (x$_1$, x$_2$, y$_1$, y$_2$)) = ((x$_2$ - x$_1$) * (y$_2$ - y$_1$) + 1) {\rm div} 2
\end{tabbing}
}

Only the first part of this lemma (the number of green squares) needed
to be explicitly proved, while the number of yellow squares is easily
calculated as the difference between the total number of squares and
the number of green squares. An analogous lemma for a yellow starting
square $(x_1, y_1)$ is proved (again not directly, but by reducing it
to the present lemma for a green starting square, by translating the
whole rectangle by one square to the right).

The correctness of the formula for the number of green squares of a
rectangle that starts on a green square is proved by induction on the
height of a rectangle ($y_2 - y_1 - 1$). Both in the base case and in
the inductive step, a lemma that characterizes the number of green
squares in a single row is used (it covers both the case of a green
and a yellow starting square in that row).

{\it
\begin{tabbing}
  \hspace{5mm}\=\hspace{5mm}\=\hspace{5mm}\=\hspace{5mm}\=\kill
{\bf lemma}\\
\>  {\bf assumes} "x1 $<$ x2"\\
\>  {\bf shows} "\=card $\{$(x, y). x1 $\leq$ x $\wedge$ x $<$ x2 $\wedge$ y = y0 $\wedge$ green (x, y)$\}$ = \\
\>\>         ({\rm if} green (x1, y0) {\rm then} (x2 - x1 + 1) {\rm div} 2 {\rm else} (x2 - x1) {\rm div} 2)"
\end{tabbing}
}

\noindent This lemma is also proved by mathematical induction, this
time over the length of the row ($x_2 - x_1 - 1$). In both inductive
proofs the set of squares of the rectangle is given as a disjoint
union of squares of a smaller rectangle (for which we know the number
of green squares by induction hypothesis) and a degenerated rectangle
(a row, i.e., a single square) for which we directly calculate the
number of green squares (depending on the color of its first square).
  
With those lemmas in place, the number of green and yellow squares in
a green rectangle can easily be connected, by also noting that both
dimensions of that rectangle must be odd, so its total number of
squares is odd. The case of a yellow rectangle is fully analogous,
while the mixed rectangle must have an even number of squares that is
evenly split between green and yellow squares.

The informal proof continues as follows. ''The rectangle $\mathcal{R}$
is green, so it contains more green unit squares than yellow unit
squares. Therefore, among the small retangles, at least one is
green.''

This is formalized by the following theorem (proved for any green rect).

{\it
\begin{tabbing}
  \hspace{5mm}\=\hspace{5mm}\=\hspace{5mm}\=\hspace{5mm}\=\kill
{\bf lemma}\\
\>  {\bf assumes} \="green\_rect (x$_1$, x$_2$, y$_1$, y$_2$)" "valid\_rect (x$_1$, x$_2$, y$_1$, y$_2$)"\\
\>\>"tiles rs (x$_1$, x$_2$, y$_1$, y$_2$)" "$\forall$ r $\in$ rs. valid\_rect r"\\
\>  {\bf shows} "$\exists$ r $\in$ rs. green\_rect r"
\end{tabbing}
}

The proof is by contradiction (we will show only the proof
outline). If the negation of the thesis is assumed, then all tiles are
either yellow or mixed, so, by the previous lemmas, they have less or
equal green than yellow squares.

{\it
\begin{tabbing}
  \hspace{5mm}\=\hspace{5mm}\=\hspace{5mm}\=\hspace{5mm}\=\kill
  {\bf proof} (rule ccontr)\\
  {\bf assume} "$\neg$ ?thesis"\\
  {\bf then have} "$\forall$ r $\in$ rs. yellow_rect r $\vee$ mixed_rect r" using mixed_rect_def  \\
\>  $\isacartoucheopen$proof$\isacartoucheclose$\\
  {\bf then have} "$\forall$ r $\in$ rs. card (green_squares r) $\leq$ card (yellow_squares r)" \\
\>  $\isacartoucheopen$proof$\isacartoucheclose$
\end{tabbing}
}

Therefore, the total number of green squares in the big rectangle is
less or equal to the number of yellow squares in the big rectangle,
which is contradictory to earlier lemma about green rectangles. This
proof goes as follows.

{\it
\begin{tabbing}
  \hspace{5mm}\=\hspace{5mm}\=\hspace{5mm}\=\hspace{5mm}\=\kill
  {\bf have} "card (green_squares (x$_1$, x$_2$, y$_1$, y$_2$)) $\leq$ card (yellow_squares (x$_1$, x$_2$, y$_1$, y$_2$))"\\
  {\bf proof}-\\
 \>    {\bf have} "card (green_squares (x$_1$, x$_2$, y$_1$, y$_2$)) = card ($\bigcup$ (green_squares ` rs))" $\isacartoucheopen$proof$\isacartoucheclose$\\
 \>    {\bf also have} "... = ($\sum$ r $\in$ rs. card (green_squares r))" $\isacartoucheopen$proof$\isacartoucheclose$\\
 \>    {\bf also have} "... $\leq$ ($\sum$ r $\in$ rs. card (yellow_squares r))" $\isacartoucheopen$proof$\isacartoucheclose$\\
 \>    {\bf also have} "... = card ($\bigcup$ (yellow_squares ` rs))" $\isacartoucheopen$proof$\isacartoucheclose$\\
 \>    {\bf also have} "... = card (yellow_squares (x$_1$, x$_2$, y$_1$, y$_2$))" $\isacartoucheopen$proof$\isacartoucheclose$\\
 \>    {\bf finally show} ?thesis .\\
  {\bf qed}\\
  {\bf then show} False {\bf using} `green_rect (x$_1$, x$_2$, y$_1$, y$_2$)` `valid_rect (x$_1$, x$_2$, y$_1$, y$_2$)` green_rect {\bf by} auto
\end{tabbing}
}

Showing that the cardinality of the union is the sum of cardinalities
of its members is not trivial, since it requires proving that the
union is disjoint and that all involved sets are finite. For example,
the outline of the second subproof in the previous proof is the
following.

{\it
\begin{tabbing}
  \hspace{5mm}\=\hspace{5mm}\=\hspace{5mm}\=\hspace{5mm}\=\kill
  {\bf have} "card ($\bigcup$ (green_squares ` rs)) = ($\sum$ r $\in$ rs. card (green_squares r))"\\
  {\bf proof} (rule card_UN_disjoint)\\
  \> {\bf show} "finite rs" $\isacartoucheopen$proof$\isacartoucheclose$\\
  \> {\bf show} "$\forall$ r $\in$ rs. finite (green_squares r)" {\bf by} auto\\
  \> {\bf show} "$\forall$ r1 $\in$ rs. $\forall$ r2 $\in$ rs. r1 $\neq$ r2 $\longrightarrow$ green_squares r1 $\cap$ green_squares r2 = $\{\}$" $\isacartoucheopen$proof$\isacartoucheclose$\\
  {\bf qed}
\end{tabbing}
}

To show that the tiling {\it rs} must contain a finite number of
rectangles we show that each tile must be inside the big rectangle,
and that there are only finitely many rectangles that can be inside a
given rectangle ({\it rs} is the subset of {\it
  $\{$x$_1$..x$_2$$\}$ $\times$
  $\{$x$_1$..x$_2$$\}$ $\times$ $\{$y$_1$..y$_2$$\}$
  $\times$ $\{$y$_1$..y$_2$$\}$}, which is finite). The disjointness
of sets of green squares follows from the fact that the tiles are
non-overlapping.

The informal proof finishes as follows. ''Let $\mathcal{S}$ be such a
small green rectangle, and let its distances from the sides of
$\mathcal{R}$ be $x$, $y$, $u$ and $v$. The top-left corner of
$\mathcal{R}$ and the top-left corner of $\mathcal{S}$ have the same
color, which happens if and only if $x$ and $u$ have the same
parity. Similarly, the other three green corners of $\mathcal{S}$
indicate that $x$ and $v$ have the same parity, $y$ and $u$ have the
same parity, i.e. $x$, $y$, $u$ and $v$ are all odd or all even.''

This is formalized as follows.

{\it
\begin{tabbing}
  \hspace{5mm}\=\hspace{5mm}\=\hspace{5mm}\=\hspace{5mm}\=\kill
{\bf definition} inside :: "rect $\Rightarrow$ rect $\Rightarrow$ bool" {\bf where} "inside ri ro $\longleftrightarrow$ squares ri $\subseteq$ squares ro"\\[1ex]
{\bf lemma}\\
\>  {\bf assumes} \="valid_rect (x$_1^i$, x$_2^i$, y$_1^i$, y$_2^i$)"  "green_rect (x$_1^i$, x$_2^i$, y$_1^i$, y$_2^i$)" "green_rect (x$_1^o$, x$_2^o$, y$_1^o$, y$_2^o$)"\\
\>\>          "inside (x$_1^i$, x$_2^i$, y$_1^i$, y$_2^i$) (x$_1^o$, x$_2^o$, y$_1^o$, y$_2^o$)"\\
\> {\bf shows} "l\=et ds = $\{$x$_1^i$ - x$_1^o$, x$_2^o$ - x$_2^i$, y$_1^i$ - y$_1^o$, y$_2^o$ - y$_2^i$$\}$ in ($\forall$ d $\in$ ds. even d) $\vee$ ($\forall$ d $\in$ ds. odd d)"
\end{tabbing}
}

Interestingly, this lemma can be proved almost fully
automatically. Even though its informal proof is given in more detail
than previous proofs that required much longer formal proofs.

\subsection{A Problem in Number Theory}

As an example of a problem in number theory we show the formalization
of the problem N1 from 2017\footnote{The problem statement and its
  solution are available in the official competition bulletin
  \url{http://www.imo-official.org/problems/IMO2017SL.pdf}.}. The
formalization follows the official solutions, but reveals many gaps
typical for informal proofs.

\begin{problem}[2017 N1]
  The sequence $a_0$, $a_1$, $a_2$, \ldots of positive integers satisfies

  \begin{equation*}
  a_{n+1}=
  \begin{cases}
    \sqrt{a_n,} & \mbox{if } \sqrt{a_n} \mbox{ is an integer}\\
    a_n + 3,    & \mbox{otherwise}
  \end{cases}
  \qquad
  \mbox{for every } n \geq 0
  \end{equation*}
  
  \noindent Determine all values of $a_0 > 1$ for which there is at
  least one number $a$ such that $a_n = a$ for infinitely many values
  of $n$.
\end{problem}

The answer is ''all positive multiples of 3'' and we can easily
formalize the problem statement. A slight problem is that {\it sqrt}
in Isabelle/HOL is defined only for real numbers, so to avoid using
reals we define the square root of naturals.

{\it
\begin{tabbing}
  \hspace{5mm}\=\hspace{5mm}\=\hspace{5mm}\=\hspace{5mm}\=\kill
{\bf definition} sqrt\_nat :: "nat $\Rightarrow$ nat" {\bf where}\\ 
\>  "sqrt\_nat x = (THE s. x = s * s)"\\[1ex]
{\bf theorem} IMO_2017_SL_N1:\\
\>  {\bf fixes} a :: "nat $\Rightarrow$ nat"\\
\>  {\bf assumes} "$\forall$ n. a (n + 1) = (if ($\exists$ s. a n  = s * s) then sqrt\_nat (a n) else (a n) + 3)" {\bf and} "a 0 $>$ 1"\\
\>  {\bf shows} "($\exists$ A. infinite {n. a n = A}) $\longleftrightarrow$ a 0 mod 3 = 0"
\end{tabbing}
}

The informal proof begins as follows. ''Since the value of $a_{n+1}$
only depends on the value of $a_n$, if $a_n = a_m$ for two different
indices $n$ and $m$, then the sequence is eventually periodic. So we
look for the values of $a_0$ for which the sequence is eventually
periodic.'' This is formulated by the following definition and a
series of lemmas.

{\it
  \begin{tabbing}    
    \hspace{5mm}\=\hspace{5mm}\=\hspace{5mm}\=\hspace{5mm}\=\kill
{\bf definition} eventually_periodic :: "(nat $\Rightarrow$ 'a) $\Rightarrow$ bool" {\bf where}\\
\>  "eventually_periodic a $\longleftrightarrow$ ($\exists$ p $>$ 0. $\exists$ n$_0$. $\forall$ n $\geq$ n$_0$. a (n + p) = a n)"\\[1ex]
{\bf lemma}\\
\>  {\bf fixes} a :: "nat $\Rightarrow$ 'a"\\
\>  {\bf assumes} "$\forall$ n $\geq$ n$_0$. a (n + p) = a n"\\
\>  {\bf shows} "$\forall$ k. a (n$_0$ + k * p) = a n$_0$"\\
$\isacartoucheopen$proof$\isacartoucheclose$\\[1ex]
{\bf lemma}\\
\>  {\bf fixes} a :: "nat $\Rightarrow$ 'a"\\
\>  {\bf assumes} "$\forall$ n. a (n + 1) = f (a n)" "a n$_1$ = a n$_2$"\\
\>  {\bf shows} "$\forall$ k. a (n$_1$ + k) = a (n$_2$ + k)"\\
$\isacartoucheopen$proof$\isacartoucheclose$\\[1ex]
{\bf lemma}\\
\>  {\bf fixes} a :: "nat $\Rightarrow$ 'a"\\
\>  {\bf assumes} "$\forall$ n. a (n + 1) = f (a n)" "n$_1$ < n$_2$" "a n$_1$ = a n$_2$"\\
\> {\bf shows} "eventually_periodic a"\\
$\isacartoucheopen$proof$\isacartoucheclose$\\[1ex]
{\bf lemma}\\
\>  {\bf fixes} a :: "nat $\Rightarrow$ 'a"\\
\>  {\bf assumes} "$\forall$ n. a (n + 1) = f (a n)"\\
\> {\bf shows} "($\exists$ A. infinite $\{$n. a n = A$\}$) $\longleftrightarrow$ eventually\_periodic a"\\
$\isacartoucheopen$proof$\isacartoucheclose$
\end{tabbing}
}

The first two lemmas are proved by induction (on the value $k$). The
third lemma is proved by applying the second lemma to prove that the
sequence periodically repeats after $n_2-n_1$ elements, starting on
the index $n_1$. The proof of the last lemma is split into two
directions. In the first direction, we assume that there exists an
infinite set of indexes that contains elements with the same value. In
that case, there must be two different values $n_1$ and $n_2$ such
that {\it a $n_1$ = a $n_2$}, so the sequence is eventually periodic
by the third lemma. In the second direction of the proof, we assume
that the sequence is periodic, then by the first lemma, the sequence
attains the value $a_{n_0}$ on the infinite set of indices
$n_0 + k\cdot p$.

The informal proof then continues by a series of claims and their
proofs.

\textbf{Claim 1.}
''If $a_n \equiv -1\,(\mathrm{mod}\; 3)$, then, for all $m > n$,
$a_m$ is not a perfect square. It follows that the sequence is
eventually strictly increasing, so it is not eventually periodic.

\noindent \emph{Proof.} A square cannot be congruent to $-1$ modulo 3,
  so $a_n \equiv -1\,(\mathrm{mod}\; 3)$ implies that $a_n$ is not a
  square, therefore $a_{n+1} = a_n + 3 > a_n$. As a consequence,
  $a_{n+1} \equiv a_n \equiv -1\,(\mathrm{mod}\; 3)$, so $a_{n+1}$ is
  not a square either. By repeating the argument, we prove that, from
  $a_n$ on, all terms of the sequence are not perfect squares and are
  greater than their predecessors, which completes the proof.''

  First we formalize the notion of being eventually increasing, give
  its equivalent characterization (proved by induction) and prove that
  a strictly increasing sequence cannot be periodic (the proof by
  contradiction is very simple since eventually strictly increasing
  and periodic sequence would have two values $a_n$ and $a_{n+p}$ for
  which it would have to hold both $a_n = a_{n+p}$ and
  $a_n < a_{n+p}$).

{\it
  \begin{tabbing}    
    \hspace{5mm}\=\hspace{5mm}\=\hspace{5mm}\=\hspace{5mm}\=\kill
{\bf definition} eventually_increasing :: "(nat $\Rightarrow$ nat) $\Rightarrow$ bool" {\bf where}\\
\>  "eventually_increasing a $\longleftrightarrow$ ($\exists$ n$_0$. $\forall$ n $\geq$ n$_0$. a n $<$ a (n + 1))"\\[1ex]
{\bf lemma}\\
\>  {\bf shows} "eventually_increasing a $\longleftrightarrow$ ($\exists$ n$_0$. $\forall$ i j. n$_0$ $\leq$ i $\wedge$ i $<$ j $\longrightarrow$ a i $<$ a j)"\\
$\isacartoucheopen$proof$\isacartoucheclose$\\[1ex]
{\bf lemma}\\
\>  {\bf assumes} "eventually_increasing a"\\
\>  {\bf shows} "$\neg$ eventually_periodic a"\\
$\isacartoucheopen$proof$\isacartoucheclose$
\end{tabbing}
}

We need to formulate and prove that ''A square cannot be congruent to
-1 modulo 3''. There is no need to use negative numbers since
$a_n\equiv -1\,(\mathrm{mod}\; 3)$ is equivalent to
$a_n\equiv 2\,(\mathrm{mod}\; 3)$.

{\it
  \begin{tabbing}
    \hspace{5mm}\=\hspace{5mm}\=\hspace{5mm}\=\hspace{5mm}\=\kill
{\bf lemma}\\
\>  {\bf fixes} s :: nat\\
\>  {\bf shows} "(s * s) mod 3 $\neq$ 2"\\
$\isacartoucheopen$proof$\isacartoucheclose$
  \end{tabbing}    
}

\emph{Claim1} establishes several facts, but the only ''takeaway'',
i.e., the only fact that is used later in the proof is that if
$a_n\equiv -1\,(\mathrm{mod}\; 3)$, then the sequence is not eventually
periodic. Since \emph{Claim1} is useful only for the proof of the main
theorem, we do not formulate it as a general lemma, but instead we
formulate it as a named intermediate fact within the proof of the main
theorem:

{\it
  \begin{tabbing}    
    \hspace{5mm}\=\hspace{5mm}\=\hspace{5mm}\=\hspace{5mm}\=\kill
    {\bf have} Claim1: "$\exists$ n. a n mod 3 = 2 $\Longrightarrow$ $\neg$ eventually\_periodic a"
\end{tabbing}
}

The informal proof gives a delicate connection between the fact that
elements are not full squares and that the sequence is strictly
increasing. The language construction ''by repeating the argument''
indicates that the proof is essentially based on mathematical
induction. Therefore, to prove the previous claim we prove the
following statement (by induction on the value $m - n$).

{\it
  \begin{tabbing}    
    \hspace{5mm}\=\hspace{5mm}\=\hspace{5mm}\=\hspace{5mm}\=\kill
    {\bf have} "∀ m $\geq$ n. ($\nexists$ s. a m = s * s) $\wedge$ a m mod 3 = 2 $\wedge$ a (m + 1) = a m + 3" $\isacartoucheopen$proof$\isacartoucheclose$
\end{tabbing}
}

From this it easily follows that the sequence is eventually
increasing, and therefore, by a previous lemma, not eventually
periodic.

\bigskip

The informal proof continues with a second claim. 

\textbf{Claim 2.}
''If $a_n \not\equiv -1\,(\mathrm{mod}\; 3)$ and $a_n > 9$ then there
is an index $m > n$ such that $a_m < a_n$.

\noindent \emph{Proof.} Let $t^2$ be the largest perfect square which
is less than $a_n$. Since $a_n > 9$, t is at least 3. The first square
in the sequence $a_n, a_n + 3, a_n + 6, \ldots$ will be $(t+1)^2,
(t+2)^2, (t+3)^2$, therefore there is an index $m > n$ such that $a_m
\leq t + 3 < t^2 < a_n$, as claimed.''

This claim is very easily stated (again as a named intermediate fact
within the main proof).

{\it
  \begin{tabbing}    
    \hspace{5mm}\=\hspace{5mm}\=\hspace{5mm}\=\hspace{5mm}\=\kill
    {\bf have} Claim2: "$\forall$ n. a n mod 3 $\neq$ 2 $\wedge$ a n $>$ 9 ⟶ ($\exists$ m $>$ n. a m $<$ a n)"
  \end{tabbing}    
}

\noindent However, formal proof of this claim is very involved, since
the informal proof is very imprecise. First we define the value of
$?t$, and prove its basic properties. It requires showing that the set
$?T$ of all perfect squares less than {\it a n} is finite and
non-empty (it contains the number 3 so it must be non-empty).

{\it
  \begin{tabbing}
    \hspace{5mm}\=\hspace{5mm}\=\hspace{5mm}\=\hspace{5mm}\=\hspace{5mm}\=\kill
    {\bf let} ?T = "$\{$t | t. t*t $<$ a n$\}$" {\bf and} ?t = "Max ?T"\\
    {\bf have} "?t $\geq$ 3" "?t$^2$ $<$ a n" "a n $\leq$ (?t + 1)$^2$" $\isacartoucheopen$proof$\isacartoucheclose$    
  \end{tabbing}    
}

The claim ''The first square in the sequence $a_n, a_n + 3, a_n + 6,
\ldots$ will be $(t+1)^2, (t+2)^2, (t+3)^2$'' was very hard to prove
formally. The statement is formalized as follows.

{\it
  \begin{tabbing}    
    {\bf have} "$\exists$ k. a (n + k) $\in$ $\{$(?t+1)$^2$, (?t+2)$^2$, (?t+3)$^2$$\}$"
  \end{tabbing}    
}

\noindent Note that we used $a_{n+k}$ instead of $a_n +
3k$. Although that is essentially the same, since these two sequences
coincide until a perfect square occurs, the sequences $a_n, a_n + 3,
a_n + 6, \ldots$ and $a_n, a_{n+1}, a_{n+2},
\ldots$ must be formally linked (and this is going to be done within
the proof). To prove the given statement, we first prove the following
number-theoretic fact.

{\it
  \begin{tabbing}    
    {\bf have} "a n mod 3 = (?t+1)$^2$ mod 3 $\vee$ a n mod 3 = (?t+2)$^2$ mod 3 $\vee$ a n mod 3 = (?t+3)$^2$ mod 3"
  \end{tabbing}    
}

\noindent Since it is assumed that {\it a n mod 3} is not 2, it must
be either 0 or 1. The proof of the fact then follows from the next
general lemma (whose proof goes by a case analysis of the value of
{\it t mod 3}).

{\it
  \begin{tabbing}
    \hspace{5mm}\=\hspace{5mm}\=\hspace{5mm}\=\hspace{5mm}\=\kill
{\bf lemma}\\
\>  {\bf fixes} t :: nat\\
\>  {\bf shows} "$\{$(t + 1)$^2$ mod 3, (t + 2)$^2$ mod 3, (t + 3)$^2$ mod 3$\}$ = $\{$0, 1$\}$"\\
$\isacartoucheopen$proof$\isacartoucheclose$
  \end{tabbing}    
}

\noindent The sought value {\it a (n+k)} that belongs to the set {\it
  $\{$(?t+1)$^2$, (?t+2)$^2$,
  (?t+3)$^2$$\}$} will be equal to the first element of the sequence
{\it (?t+1)$^2$, (?t+2)$^3$,
  (?t+3)$^2$} that is congruent to {\it a n} modulo 3. We prove this
in the form of the following auxiliary claim.

{\it
  \begin{tabbing}
    \hspace{5mm}\=\hspace{5mm}\=\hspace{5mm}\=\hspace{5mm}\=\kill
\>    {\bf fix} i\\
\>    {\bf assume} \="i > 0" {\bf and} "$\forall$ i'. 0 $<$ i' $\wedge$ i' $<$ i $\longrightarrow$ a n mod 3 $\neq$ (?t + i')$^2$ mod 3" {\bf and}\\
\>\> "a n mod 3 = (?t + i)$^2$ mod 3"\\
\>{\bf have} "$\exists$ k. a (n + k) = (?t + i)$^2$"\\
\>$\isacartoucheopen$proof$\isacartoucheclose$
  \end{tabbing}
}

\noindent The sought index $?k$ is {\it ((?t + i)$^2$ - a n) div 3}.
We prove that {\it a$_{n + ?k}$} will be equal to {\it (?t + i)$^2$},
and that it will be the first perfect square in the sequence {\it
  a$_n$, a$_{n+1}$, \ldots} This follows from the following fact:

{\it
  \begin{tabbing}
    \hspace{5mm}\=\hspace{5mm}\=\hspace{5mm}\=\hspace{5mm}\=\hspace{5mm}\=\kill
    {\bf have} "$\forall$ k' $\leq$ ?k. a (n + k') = a n + 3 * k'"
    \hspace{5mm}
  \end{tabbing}
}

\noindent The proof goes by induction on $k'$. The base case is
trivial. In the inductive step we need to show that the equality holds
for $k'+1 \leq ?k$ under the assumption that it holds for $k' <
?k$. It suffices to prove that \mbox{{\it a (n+k')}} is not a full
square. We prove that by contradiction. If it were a full square,
since {\it k'+1 $\leq$ ?k = ((?t + i)$^2$ - a n) div 3} it would hold
that {\it 3 * (k' + 1) $\leq$ (?t + i)$^2$ - a n}, i.e., that {\it a
  (n+k') = a n + 3 * k' $\leq$ (?t + i)$^2$ - 3}. Therefore,
$a (n+k')$ would be a full square strictly between $?t^2$ and
$(?t+i)^2$, which is impossible by our assumption that {\it $\forall$
  i'. 0 $<$ i' $\wedge$ i' $<$ i $\longrightarrow$ a n mod 3 $\neq$
  (?t + i')$^2$ mod 3}. Since \mbox{{\it a (n+k')}} is not a full
square, it holds that {\it a (n+k'+1) = a (n+k') + 3 = a n + 3*k + 3 =
  a n + 3*(k+1)}.  This finishes the inductive proof.

\noindent Therefore, {\it a (n + ?k) = a (n + 3*?k)} and {\it a (n +
  3*?k) = (?t + i)$^2$} (this holds directly by the definition of {\it
  ?k = ((?t + i)$^2$ - a n) div 3}). So there indeed exists $k$ such
that {\it a (n + k) = (?t + i)$^2$}, which finishes the proof of the
axuiliary claim.

\noindent The statement {\it $\exists$ k. a (n + k) $\in$
  $\{$(?t+1)$^2$, (?t+2)$^2$,
  (?t+3)$^2$$\}$} is then proved by case analysis of the fact {\it a
  n mod 3 = (?t+1)$^2$ mod 3 $\vee$ a n mod 3 =
  (?t+2)$^2$ mod 3 $\vee$ a n mod 3 =
  (?t+3)$^2$ mod 3}, applying the auxiliary claim for $i=1$,
$i=2$, and $i=3$. From that it easily follows that {\it
  $\exists$ k. a (n + k + 1) $\in$ $\{$?t+1, ?t+2,
  ?t+3$\}$}, so {\it a (n + k + 1) $\leq$ ?t + 3 $<$ ?t$^2$
  $<$ a n}, finishing he formal proof of \emph{Claim2}.

\bigskip

The next claim in the informal proof is the following.

\textbf{Claim 3.}
''If $a_n \equiv 0\ (\mathrm{mod}\; 3)$, then there is an
index $m > n$ such that $a_m = 3$.

\noindent \emph{Proof.} First we notice that, by the definition of the
sequence, a multiple of 3 is always followed by another multiple of
3. If $a_n \in \{3, 6, 9\}$ the sequence will eventually follow the
periodic pattern $3, 6, 9, 3, 6, 9, \ldots$. If $a_n > 9$, let $j$ be
an index such that $a_j$ is equal to the minimum value of the set
$\{a_{n+1}, a_{n+2}, \ldots\}$. We must have $a_j \leq 9$, otherwise
we could apply \emph{Claim2} to $a_j$ and get a contradiction on the
minimality hypothesis. It follows that $a_j \in \{3, 6, 9\}$, and the
proof is complete.''

By analyzing the informal proof we note that the result for the case
when $a_n \leq 9$ is also applied within the case $a_n >
9$. Therefore, it is wise to prove that case as the following
sub-claim.

{\it
  \begin{tabbing}
    \hspace{5mm}\=\hspace{5mm}\=\hspace{5mm}\=\hspace{5mm}\=\kill
  {\bf have} Claim3_a: "$\forall$ n. a n mod 3 = 0 $\wedge$ a n $\leq$ 9 ⟶ ($\exists$ m $>$ n. a m = 3)"
 \end{tabbing}
}

\noindent It is easy to prove by using the recursive definition of
{\it a} and direct calculations, except one little detail. If {\it a n
  mod 3 = 0}, and {\it a n $\leq$ 9} then {\it n} could be 0, 3, 6, or
9. In the case 3, 6, and 9, the claim then easily follows by direct
calculations using the definition of the sequence {\it a}. This is not
the case if {\it a n = 0}. To prove that this case is impossible, we
must again use induction to prove.

{\it
  \begin{tabbing}
    \hspace{5mm}\=\hspace{5mm}\=\hspace{5mm}\=\hspace{5mm}\=\kill
    {\bf have} "$\forall$ n. a n$ >$ 1"
 \end{tabbing}
}

\noindent This proof uses the hypothesis {\it a 0 > 1}, and the fact
that the square root of a number that is strictly greater than 1 must
itself be a number strictly greater than 1.

{\emph Claim3} is then stated as follows:

{\it
  \begin{tabbing}
    \hspace{5mm}\=\hspace{5mm}\=\hspace{5mm}\=\hspace{5mm}\=\kill
  {\bf have} Claim3: "$\forall$ n. a n mod 3 = 0 ⟶ ($\exists$ m $>$ n. a m = 3)"
 \end{tabbing}
}

Its proof starts by case analysis on whether {\it a n $\leq$ 9}. The
case when {\it a n $\leq$ 9} is already covered by the sub-claim. If
{\it a n > 9}, we follow the informal proof by taking the minimal
element of the set \mbox{{\it a ` $\{$n+1..$\}$}} (it is the image of
the set {\it $\{$n+1..$\}$} under the function {\it
  a}). Unfortunately, we cannot use the operator {\it Min} to find its
minimum, since it is intended only for finite sets. An arbitrary
infinite set does not need to have a minimal element. However, this is
the set of natural numbers, and one of the characteristic features of
natural numbers is that they are well-ordered i.e., that each
non-empty set contains a minimal element. Isabelle/HOL supports the
binder {\it LEAST} that determines the least element in a well-ordered
set and we will use it to define the value $?m$.

{\it
  \begin{tabbing}
    \hspace{5mm}\=\hspace{5mm}\=\hspace{5mm}\=\hspace{5mm}\=\kill
    {\bf let} ?m = "LEAST x. x $\in$ (a ` $\{$n+1..$\}$)"\\
    {\bf let} ?j = "SOME j. j $>$ n $\wedge$ a j = ?m"
 \end{tabbing}
}

\noindent Note the use of the indefinite description operator {\it
  SOME} in the definition of {\it ?j}. This is justified by the fact
that {\it ?m} is a member of {\it a ` $\{$n+1..$\}$}. The proof
continues by case analysis on {\it a ?j $\leq$ 9}. If that holds, then
\emph{Claim3} follows by \emph{Claim3_a}. Otherwise we apply
\emph{Claim2}. To use it we must establish that {\it a ?j mod 3 $\neq$
  2}. This follows from the first sentence in the informal proof:
''First we notice that, by the definition of the sequence, a multiple
of 3 is always followed by another multiple of 3''. We formalize this
by the following statement:

{\it
  \begin{tabbing}
    \hspace{5mm}\=\hspace{5mm}\=\hspace{5mm}\=\hspace{5mm}\=\kill
    {\bf have} "$\forall$ n n'. a n mod 3 = 0 $\wedge$ n $\leq$ n' $\longrightarrow$ a n' mod 3 = 0"
 \end{tabbing}
}

\noindent Now we use induction on {\it n' - n} in the proof.  This
proof also requires proving and using the following simple
number-theoretic lemma.

{\it
  \begin{tabbing}
    \hspace{5mm}\=\hspace{5mm}\=\hspace{5mm}\=\hspace{5mm}\=\kill
{\bf lemma}\\
\>  {\bf fixes} x :: nat\\
\>  {\bf shows} "(x * x) mod 3 = 0 ⟷ x mod 3 = 0"\\
\>$\isacartoucheopen$proof$\isacartoucheclose$
 \end{tabbing}
}

\noindent Since {\it a n mod 3 = 0}, it must hold that {\it a ?j mod 3
  = 0 $\neq$ 2}, so \emph{Claim2} can be used to obtain {\it m $>$ ?j}
such that {\it a m $<$ a ?j}, but this clearly contradicts the
definition of {\it ?j}.

\bigskip

Finally, the informal proof gives the last claim.

\textbf{Claim 4.}
''If $a_n \equiv 1\ (\mathrm{mod}\; 3)$, then there is an index $m > n$
such that $a_m \equiv -1\ (\mathrm{mod}\; 3)$.

\noindent \emph{Proof.} In the sequence, 4 is always followed by
$2 \equiv -1\ (\mathrm{mod}\; 3)$, so the claim is true for $a_n = 4$.
If $a_n = 7$, the next terms will be 10, 13, 16, 4, 2, \ldots and the
claim is also true. For $a_n \geq 10$, we again take an index $j > n$
such that $a_j$ is equal to the minimum value of the set
$\{a_{n+1}, a_{n+2}, \ldots \}$, which by the definition of the
sequence consists of non-multiples of 3. Suppose
$a_j = 1\ (\mathrm{mod}\; 3)$.  Then we must have $a_j \leq 9$ by
\emph{Claim2} and the minimality of $a_j$. It follows that
$a_j \in \{4, 7\}$, so $a_m = 2 < a_j$ for some $m > j$, contradicting
the minimality of $a_j$. Therefore, we must have
$a_j \equiv -1\ (\mathrm{mod}\; 3)$.''

\medskip

Similar as we did for the \emph{Claim3}, first we prove the following
sub-claim.

{\it
  \begin{tabbing}
    \hspace{5mm}\=\hspace{5mm}\=\hspace{5mm}\=\hspace{5mm}\=\kill
    {\bf have} Claim4_a: "$\forall$ n. a n mod 3 = 1 $\wedge$ a n $\leq$ 9 ⟶ ($\exists$ m > n. a m mod 3 = 2)"
 \end{tabbing}
}

\noindent The proof is again using direct calculations, the recursive
definition of the sequence $a$ and earlier established fact that {\it
  $\forall$ n. a n > 1}.

\emph{Claim4} is then formulated as follows.

{\it
  \begin{tabbing}
    \hspace{5mm}\=\hspace{5mm}\=\hspace{5mm}\=\hspace{5mm}\=\kill
    {\bf have} Claim4: "$\forall$ n. a n mod 3 = 1 ⟶ ($\exists$ m > n. a m mod 3 = 2)"
 \end{tabbing}
}

\noindent The formal proof of this claim is very similar to the
formalization of \emph{Claim3}, so we omit the details. Formalizing
the sentence ''the set $\{a_{n+1}, a_{n+2}, \ldots \}$, which by the
definition of the sequence consists of non-multiples of 3'', required
proving the following fact (by induction on {\it n' - n}).

{\it
  \begin{tabbing}
    \hspace{5mm}\=\hspace{5mm}\=\hspace{5mm}\=\hspace{5mm}\=\kill
    {\bf have} "$\forall$ n n'. a n mod 3 $\neq$ 0 $\wedge$ n $\leq$ n' $\longrightarrow$ a n' mod 3 $\neq$ 0"
 \end{tabbing}
}

\bigskip

Once all four claims are formally proved, the theorem itself can be
proved. Informal proof goes as follows.

\textbf{Main theorem.}
''It follows from the previous claims that if $a_0$ is a multiple of 3
the sequence will eventually reach the periodic pattern $3, 6, 9, 3,
6, 9, \ldots$; if $a_0 \equiv -1\ (\mathrm{mod}\; 3)$ the sequence will
be strictly increasing; and if $a_0 \equiv 1\ (\mathrm{mod}\; 3)$ the
sequence will eventually be strictly increasing. So the sequence will
eventually be periodic if and only if, $a_0$ is a multiple of 3.''

We show the full formal proof of this part (note that \emph{Claim3} is
used in the first direction, while the second direction uses only
\emph{Claim1} and \emph{Claim4}, while \emph{Claim2} is only a lemma
used with the proof of \emph{Claim3} and \emph{Claim4}, and not in the
main proof).

{\it
  \begin{tabbing}
    \hspace{5mm}\=\hspace{5mm}\=\hspace{5mm}\=\hspace{5mm}\=\kill
  {\bf show} ?thesis\\
  {\bf proof}\\
\>    {\bf assume} "a 0 mod 3 = 0"\\
\>    {\bf then have} "eventually_periodic a" {\bf using} Claim3 two_same_periodic[OF assms(1)] {\bf by} simp\\
\>    {\bf then show} "$\exists$ A. infinite $\{$n. a n = A$\}$" {\bf using} infinite_periodic[OF assms(1)] {\bf by} simp\\
  {\bf next}\\
\>    {\bf assume} "$\exists$ A. infinite $\{$n. a n = A$\}$"\\
\>    {\bf then have} "eventually\_periodic a" {\bf using} infinite_periodic[OF assms(1)] {\bf by} simp\\
\>    $\{$\\
\>\>      {\bf assume} "a 0 mod 3 = 1"\\
\>\>      {\bf then obtain} m where "a m mod 3 = 2" {\bf using} Claim4 {\bf by} auto\\
\>\>      {\bf then have} False {\bf using} Claim1 $\isacartoucheopen$eventually_periodic a$\isacartoucheclose$ {\bf by} force\\
\>    $\}$\\
\>    {\bf moreover}\\
\>    $\{$\\
\>\>      {\bf assume} "a 0 mod 3 = 2"\\
\>\>      {\bf then have} False {\bf using} Claim1 $\isacartoucheopen$eventually_periodic a$\isacartoucheclose$ {\bf by} force\\
\>    $\}$\\
\>    {\bf ultimately} {\bf show} "a 0 mod 3 = 0" {\bf by} presburger\\
  {\bf qed}
 \end{tabbing}
}

%{\it
%  \begin{tabbing}
%    \hspace{5mm}\=\hspace{5mm}\=\hspace{5mm}\=\hspace{5mm}\=\kill
% \end{tabbing}
%}

\section{Educational aspects}

The repository of formalized IMO problems is created with an ITP
course at Faculty of Mathematics, University of Belgrade. It is an
elective course at the fourth, final year of undergraduate
studies. All students enrolled already passed courses in mathematical
logic, functional programming, and many other classic mathematical
topics during their previous studies (algebra, analysis,
combinatorics, numerical mathematics etc.), so they already have quite
a good understanding of all main concepts underlying ITP. The course
is taught 14 weeks with 2 hours of lecture and 3 hours of exercises
per week, and covers formalization of mathematics and elements of
software verification in Isabelle/HOL. Grading is done by several
on-site tests with small and simple problems, and a larger project,
done off-site (at home), scoring up to 40\% of the total
points. Formalizing an IMO problem or a similar problem given at
Serbian national level competitions is given as one possible project
assignment (students can also verify some elementary algorithm or
formalize several theorems from some introductory mathematics course).

Each student selects a problem and formalizes it. No collaboration
between students is allowed, but when students get stuck, they can get
help from the teachers ``free of charge'' (they do not lose any
points for asking help). The course is new and has been given only
twice, but our experience shows that most students manage to finish
their project (some on their own, and some after several rounds of
guidance by the teachers). Of course, it is very important that
students are given informal proofs in advance, since IMO problems,
although very elementary, are very challenging and hard to
solve. Developed formal proofs are usually not the most elegant ones,
as students often introduce definitions and lemmas that already exist
in the vast Isabelle/HOL libraries and often give long, manual Isar
proofs for statements that could be proved automatically if advanced
automated proof methods are setup correctly. They are not penalized
for this, but are sometimes required to ''polish'' their proofs by
following detailed guidelines given by the teachers.

Formalizing IMO problem solutions is a very good task for practicing
interactive theorem proving and we advocate that they should be used
in courses of formal theorem proving.

\begin{itemize}
\item Although they are very hard, problems are usually formulated in
  elementary terms of high-school mathematics and do not require any
  knowledge of advanced mathematical concepts. Therefore, all students
  of mathematics and computer science can easily understand
  them. Official IMO solutions do not use any advanced theorems and
  proof steps are justified by using elementary statements that are
  already available in most proof assistants.
\item Formalizing a problem solution usually requires several
  hours. Isar proofs are usually around several hundred lines of code
  (the shortest proof we formalized was 95 LOC, while the longest was
  2024 LOC, although that depends on the code indenting
  style). Therefore, such problems should not be used in limited-time
  on-site exams, but they are perfect for homework and project
  assignments.
\item A rich repository of manually formalized solutions might offer
  a good ground for developing and training methods for automated
  solving of IMO problems (and hopefully contribute to the IMO grand
  challenge).
\end{itemize}

Although IMO competitors are high-school pupils, we do not yet have
any experience in formalizing IMO solutions with that population. We
suppose that teaching use of proof assistants would be too
demanding. On the other hand, we think that analyzing existing problem
formalizations could help the most advanced competitors in recognizing
the subtlest proof details and mastering the highest level of
mathematical precision and rigor, that could help reaching maximal
scores in competitions.

\section{Related Work}

Relationship between informal proofs and their formal counterparts is
often discussed in the literature. One example, is Dana Scott's
Foreword of the Freek Wiedijk's seminal paper comparing 17 theorem
provers~\cite{seventeen-provers}. Scott showed examples of proofs done
by changing problem representation (e.g., algebraic and geometric) and
showed examples of proofs that involve augmenting the original problem
with supporting elements (e.g., auxiliary points and lines introduced
in a geometric configuration), that make the original problem
significantly easier to solve. Scott argued that although these
changes are often easily realized and understood by humans,
computers on the other hand have much difficulty in finding such
proofs. A popular informal proof method is given by the so-called
''Proofs Without Words'', where the property is intuitively described
by a convenient figure (e.g., there are many such diagrams that
illustrate the Pythagorean theorem). One famous problem with such a
proof is the calisson
puzzle\footnote{\url{http://nau.edu/wp-content/uploads/sites/145/NAU-High-School-Math-Day-The-Calissons-Problem-fall-19.pdf}}.
Although ''the proof'' is very intuitive (and indeed remarkable),
E.~W.~Dijkstra criticized that it is an example of ''an elaborate
nonproof'' (since key arguments are not formally stated nor proved, a
slight change of the problem could yield ''the proof'' incorrect).

Although the body of formalized mathematics is growing every day,
examples of formalized IMO solutions or similar types of problems
are scarce: Manuel Eberl has formalized three out of six problems from
IMO 2019 (Q1, Q4, and Q5) \cite{IMO2019-AFP}, and several problems
statements have been formally encoded in Lean
(\url{http://github.com/IMO-grand-challenge/formal-encoding}).

One library of formalized solutions of high-school problems is
presented by Pham, Bertot and Narbox~\cite{geocoq}. They developed a
dynamic geometry proving tool for interactive proving for high-school
students and used a specific axiomatic system adapted to this task
using the notion of vectors. Sana Stojanović-Đurđević (the second
author of the current paper), proposed a method for proving
high-school problems in geometry by using coherent logic and a set of
automated theorem provers to fill-in the gaps in the manually
generated proof outlines~\cite{informaltoformal}. They used a
semi-formal TPTP-like language. The method has been successfully
applied to a collection of geometric problems from Serbian high-school
textbooks in geometry. Generated proofs are automatically translated
to Coq and Isabelle/HOL and formally verified.

\section{Conclusions and Further Work}

In this paper we have presented Isabelle/HOL formalization of several
official IMO problem solutions. The formalization is created within
the Interactive Theorem Proving course on Faculty of Mathematics,
University of Belgrade, and is available in a repository
\url{http://github.com/filipmaric/IMO}. Our experience shows that most
final year undergraduate students at the end of the course can
successfully cope with such assignments, if they are given enough
time, guidance and support by the teachers.

Our experience shows that the difference between formal and informal
proof significantly varies, mainly depending on the category of the
problem. Problems in algebra usually have very rigorous informal
proofs, that are easy to formalize. Unlike those, problems in
combinatorics usually give a very rough proof outline, that requires
significant effort to formalize. In many cases a significant effort is
required even to give a precise problem statement (problems in
combinatorics usually require many introductory definitions). The
proofs in number theory are somewhere in between (depending on the
problem).

If human competitors were to generate formal proofs, the competition
would be much harder for them (and definitely unsuitable for
high-school pupils). Our analysis shows that many official solutions
(that would certainly be graded by maximal scores) are quite far from
fully formal and often resort to obviousness and intuition. Therefore,
in our opinion, the current formulation of the IMO grand challenge is
extremely unfair for the artificial intelligence. We suggest to split
the challenge in two parts:

\begin{itemize}
\item Informal challenge, that would require automated provers only to
  generate high-level proof outlines, that can be manually judged, the
  same way as pupils solutions are judged.
\item Formal challenge, that would require provers to generate machine
  checkable proofs given high-level proof outlines of various
  granularity (the coarsest one being only the formal problem
  statement).
\end{itemize}

Our repository is open and we hope that a wider community of
contributors will be formed. Everyone is invited to contribute either
by formalizing new IMO problem statements, or by providing alternative
solutions to existing problems. We assume that many existing formal
proofs could be shortened and better automated and we invite
contributors to provide such proofs.

\bibliographystyle{eptcs}
\bibliography{IMO.bib}

\end{document}